\documentclass[aps,pra,showpacs,twocolumn,floatfix]{revtex4}
\usepackage{graphicx}% Include figure files
\usepackage{dcolumn}% Align table columns on decimal point

\begin{document}

\title{Relativistic many-body calculations of excitation energies
 and transition rates from core-excited states
in copperlike ions}

\author{U. I. Safronova and W. R. Johnson}
 \affiliation{ Department of Physics, University of Notre Dame, Notre Dame, IN 46566}

\author{A. Shlyaptseva and S. Hamasha}
 \affiliation{Physics Department/220, University of Nevada, Reno, NV
 89557}

\date{\today}% It is always \today, today,

\begin{abstract}
Energies of $(3s^23p^63d^94l4l')$, $(3s^23p^53d^{10}4l4l')$, and
$(3s3p^63d^{10}4l4l')$ states for Cu-like ions with $Z$ = 30 -100
are evaluated to second order in relativistic many-body
perturbation theory (RMBPT) starting from a Ni-like  Dirac-Fock potential.
Second-order Coulomb and Breit-Coulomb interactions are included.
Correction for the frequency-dependence of the  Breit interaction
is taken into account in lowest order. The Lamb shift correction
to energies is also included in lowest order. Intrinsic
particle-particle-hole contributions to  energies are found to be
20-30\% of the sum of  one- and two-body contributions. Transition rates and line
strengths are calculated for the $ 3l-4l'$
  electric-dipole (E1)
transitions in Cu-like ions with nuclear charge $Z$ = 30 - 100.
RMBPT including the Breit interaction is used to evaluate retarded
E1 matrix elements in length and velocity forms.  First-order
RMBPT is used to obtain intermediate coupling coefficients and
second-order RMBPT is used to calculate transition matrix
elements.  A detailed discussion of the various contributions to
the dipole matrix elements and energy levels is given for
copperlike tungsten ($Z$ = 74). The transition energies used in
the calculation of oscillator strengths and transition rates are
from second-order RMBPT. Trends of the transition rates as
functions of $Z$ are illustrated graphically for selected
transitions.
 Comparisons are
made with available experimental data.
These atomic data are important in modeling of M-shell
radiation spectra of heavy ions generated in electron beam ion trap
experiments and in M-shell diagnostics of plasmas.

\end{abstract}

\pacs{31.15.Ar, 31.15.Md, 31.25.Jf, 32.30.Rj}

%31.15.Ar  Ab-initio Calculations
%31.15.Md  Perturbation Theory
%31.25.Jf  Electron correlation calculations for atoms and ions: excited states
%32.30.Rj  X-ray spectra

%\keywords{Suggested keywords}%Use showkeys class option if
%keyword display desired

\maketitle

\section{Introduction}
This work further develops the application of the relativistic
 many-body perturbation theory (RMBPT) to the studies of atomic characteristics of
particle-particle-hole excitations of closed-shell ions.
 Recently, RMBPT calculations of energies \cite{na-en} and transition rates
 \cite{na-tr} in Na-like ions have been performed.
 The present paper focuses on the RMBPT calculations of energies and
 transition rates in Cu-like ions.
These transitions form satellite lines to the brightest Ni-like ions and
are of a great importance for M-shell diagnostics of heavy ions.

 The second-order RMBPT calculations for Cu-like ions start from a
$1s^22s^22p^63s^23p^63d^{10}$ [Ni] Dirac-Fock potential.
All possible $3l$ holes and $4l4l'$ particles leading to 563
odd-parity and 571 even-parity $3l^{-1}4l'4l"(J)$ states are considered.
 The energies of the 1134 states and transition probabilities
of the 2294 electric-dipole lines are calculated for Cu-like ions
with $Z$ = 30-100. Transitions from the  $(3s^23p^63d^94l4l')$,
$(3s^23p^53d^{10}4l4l')$, and $(3s3p^63d^{10}4l4l')$ states to the
ground ($3s^23p^63d^{10}4s$) or singly excited
($3s^23p^63d^{10}4l$) states (with $l=p, d, f$)  form satellite
lines to the bright electric-dipole (E1) lines in Ni-like ions.
These core-excited states
 (or often called doubly-excited states) in copperlike ions have been
studied extensively both experimentally and theoretically in the
past 15-20 years.

Experimentally, these configurations
have been studied by photon and electron emission spectroscopy.
 To our knowledge, the first measurements of $3d-4p$ transitions in Cu-like W and Tm
 were done by  ~\citet{cu1} by classification of
 x-ray spectra from laser
produced plasmas in the range 6 - 9 \AA. It was shown
 that  most of the Cu-like $3d-4p$ line radiation
came from    $(3s^23p^63d^{10}4s-3s^23p^63d^94s4p)$,
$(3s^23p^63d^{10}4p-3s^23p^63d^94p^2)$,
$(3s^23p^63d^{10}4d-3s^23p^63d^94p4d)$, and
$(3s^23p^63d^{10}4f-3s^23p^63d^94p4f)$ transitions.
 Wavelengths and transition probabilities were calculated in ~\cite{cu1}
 by the relativistic parametric potential method ~\cite{RPPM}.
The same method was extended by ~\citet{klapisch} to study x-ray
spectra from laser produced plasmas of atoms from Tm ($Z$ = 69) up
to Pt ($Z$ = 78). An extended analysis of x-ray spectra of laser
produced gold plasmas has been performed by ~\citet{luc} including
$3d-4f$, $3d-4p$, and $3p-4s$ transitions in Ni-, Co-, Cu-, Zn-,
and Ga-like Au. The Ni-, Co-, Cu-, Zn-, Ga-, Ge-, and As-like
isoelectronic sequences have been considered in Ref.~\cite{wyart}
to investigate the x-ray spectra of laser-irradiated elements from
tantalum ($Z$ = 73) to lead ($Z$ = 82). In addition to the above
mentioned isoelectronic sequences, the Fe- and Mn-like states have
been included by ~\citet{zigler} to analyze x-ray spectra emmited
by laser-produced plasmas of lanthanum ($Z$ = 57) and praseodymium
($Z$ = 59). The wavelengths and transition probabilities have been
calculated in ~\cite{luc,wyart,zigler}
 by the relativistic parametric potential method ~\cite{RPPM}.
 Investigation of the x-ray spectrum emitted by laser produced barium
 plasmas has been recently described by ~\citet{doron-ba,doron-pra98}.
 The RELAC relativistic atomic
code ~\cite{RPPM} has been used to identify $3d-nl$ ($n$ = 4 to 8),
$3p-4s$, and $3p-4d$ transitions of Ni-like Ba$^{28+}$ and
corresponding satellite transitions in neighboring ionization
states: Fe-, Co-, Cu-, Zn-, Ga-, and Ge-like ions.
 The X-ray spectrum emitted by a laser produced cerium plasma in the range
7.5 - 12 \AA~  has been recently investigated in detail by ~\citet{doron-pra,doron-scr}.
The RELAC computer code
~\cite{RPPM} has been used to study x-ray spectra from highly charged
tungsten ions in tokamak plasmas in the range 7 - 10 \AA~
\cite{fourner}.
%Refs.~\cite{w-pra,w-scr,recomb,china}.

In the present paper,  RMBPT is implemented to determine energies of
$(3s^23p^63d^94l4l')$, $(3s^23p^53d^{10}4l4l')$, and
$(3s3p^63d^{10}4l4l')$ states for Cu-like ions with nuclear
charges in the range of $Z$ = 30 --100. The calculations are
carried out to second order in perturbation theory and include
second-order Coulomb and Breit  interactions. Corrections for the
frequency-dependent Breit interaction are taken into account in
the lowest order. Screened self-energy and vacuum polarization
data given by  \citet{kim}  are used to
determine the QED correction.

RMBPT is used to evaluate  matrix elements, line strengths, and
transition rates for 2294 allowed and forbidden  electric-dipole
transitions between the 1332 even-parity core-excited states
[Ni]($3d^{-1}+3s^{-1})(4s^2+4s4d+4p^2+4d^2+4p4f+4f^2)$ +
[Ni]($3p^{-1})(4s4p+4s4f+4p4d+4d4f)$, and the singly excited
[Ni]($4p+4f$) states and the 962 odd-parity core-excited states
[Ni]($3p^{-1})(4s^2+4s4d+4p^2+4d^2+4p4f+4f^2)$ +
[Ni]($3d^{-1}+3s^{-1})(4s4p+4s4f+4p4d+4d4f)$ and the ground state
[Ni]($4s$) together with the singly excited  [Ni]($4d$) states in
Cu-like ions with nuclear charges ranging from $Z$ = 30 to 100.
  Retarded E1 matrix elements are
evaluated in both length and velocity forms. These calculations
start from a [Ni] Dirac-Fock potential. First-order perturbation
theory is used to obtain intermediate coupling coefficients and
second-order RMBPT is used to determine transition matrix
elements. The transition energies employed in the calculations of line
strengths and transition rates are derived from second-order
RMBPT.

\begin{table*}
\caption{\label{tab-e2}
 Second-order contributions to the energy matrices
(a.u.)\
 for odd-parity states with
$J$=1/2 in the case of Cu-like tungsten, $Z$=74. One-body,
 two-body, ant three-body second-order Coulomb  contributions
are given in columns labelled $E^{(2)}_{1}$, $E^{(2)}_{2}$, and
$E^{(2)}_{3}$, respectively.}
\begin{ruledtabular}
\begin{tabular}{llrrrr}
\multicolumn{1}{c}{$4l_{1}j_{1}4l_{2}j_{2}[J_{12}]3l_{3}$}&
\multicolumn{1}{c}{$4l_{1}j_{1}4l_{2}j_{2}[J_{12}]3l_{3}$}&
\multicolumn{1}{c}{$E^{(2)}_{1}$}&
\multicolumn{1}{c}{$E^{(2)}_{2}$}&
\multicolumn{1}{c}{$E^{(2)}_{3}$}&
\multicolumn{1}{c}{$E^{(2)}_{\rm tot}$\rule{0ex}{2.3ex}}\\
\hline
$4s_{1/2}4p_{1/2}(1)3d_{3/2}$&$4s_{1/2}4p_{1/2}(1)3d_{3/2}$&  -0.247886&  0.104563&  0.025088&  -0.118235\\
$4s_{1/2}4p_{3/2}(2)3d_{5/2}$&$4s_{1/2}4p_{3/2}(2)3d_{5/2}$&  -0.226244&  0.113517&  0.020801&  -0.091926\\
$4s_{1/2}4p_{3/2}(1)3d_{3/2}$&$4s_{1/2}4p_{3/2}(1)3d_{3/2}$&  -0.237415&  0.076978&  0.029703&  -0.130734\\
$4s_{1/2}4p_{3/2}(2)3d_{3/2}$&$4s_{1/2}4p_{3/2}(2)3d_{3/2}$&  -0.237415&  0.124305&  0.021436&  -0.091674\\
$4p_{1/2}4d_{3/2}(2)3d_{5/2}$&$4p_{1/2}4d_{3/2}(2)3d_{5/2}$&  -0.240289&  0.122424&  0.028030&  -0.089836\\
$4p_{1/2}4d_{5/2}(2)3d_{5/2}$&$4p_{1/2}4d_{5/2}(2)3d_{5/2}$&  -0.236161&  0.031594&  0.009588&  -0.194980\\
$4p_{1/2}4d_{5/2}(3)3d_{5/2}$&$4p_{1/2}4d_{5/2}(3)3d_{5/2}$&  -0.236161&  0.001311& -0.002282&  -0.237132\\
$4p_{1/2}4d_{3/2}(1)3d_{3/2}$&$4p_{1/2}4d_{3/2}(1)3d_{3/2}$&  -0.251460&  0.040039&  0.036490&  -0.174931\\
$4p_{1/2}4d_{3/2}(2)3d_{3/2}$&$4p_{1/2}4d_{3/2}(2)3d_{3/2}$&  -0.251460&  0.060163&  0.026395&  -0.164902\\
$4s_{1/2}4f_{5/2}(2)3d_{5/2}$&$4s_{1/2}4f_{5/2}(2)3d_{5/2}$&  -0.229530&  0.067699&  0.039638&  -0.122194\\
\end{tabular}
\end{ruledtabular}
\end{table*}

\begin{table*}
\caption{\label{tab-e4} Energies  of selected odd-parity levels with
$J$=1/2 of Cu-like tungsten, $Z$=74 in
a.u. $E^{(0+1)} \equiv E_{0} + E_{1} + B_{1}$. }
\begin{ruledtabular}
\begin{tabular}{lrrrrr}
\multicolumn{1}{c} {$jj$ coupling} &
 \multicolumn{1}{c}{$E^{(0+1)}$} &

 \multicolumn{1}{c}{$ B_{1} $}&
\multicolumn{1}{c}{$ E_{2} $} &
 \multicolumn{1}{c}{$E_{\rm LAMB}$}&

\multicolumn{1}{c}{$E_{\rm tot}$\rule{0ex}{2.3ex}}\\
\hline
$4s_{1/2}4p_{1/2}(1)3d_{3/2}$& -25.954298& -0.013561& -0.112483&  0.069035& -26.011307\\
$4s_{1/2}4p_{3/2}(2)3d_{5/2}$& -24.879634& -0.005093& -0.090398&  0.065104& -24.910021\\
$4s_{1/2}4p_{3/2}(1)3d_{3/2}$& -22.639268& -0.054704& -0.090313&  0.073682& -22.710603\\
$4s_{1/2}4p_{3/2}(2)3d_{3/2}$& -21.914637& -0.057899& -0.125159&  0.073518& -22.024177\\
$4p_{1/2}4d_{3/2}(2)3d_{5/2}$& -16.127594&  0.032360& -0.088446& -0.001127& -16.184807\\
$4p_{1/2}4d_{5/2}(2)3d_{5/2}$& -14.892864& -0.001171& -0.108506&  0.001320& -15.001221\\
$4p_{1/2}4d_{5/2}(3)3d_{5/2}$& -14.149853&  0.000699& -0.136464&  0.003373& -14.282246\\
$4p_{1/2}4d_{3/2}(1)3d_{3/2}$& -13.066261& -0.033638& -0.123168&  0.007346& -13.215720\\
$4p_{1/2}4d_{3/2}(2)3d_{3/2}$& -12.799288& -0.034773& -0.108188&  0.050678& -12.891570\\
$4s_{1/2}4f_{5/2}(2)3d_{5/2}$& -12.654131& -0.048963& -0.119286&  0.050671& -12.771709\\
\end{tabular}
\end{ruledtabular}
\end{table*}

\begin{table*}
\caption{Uncoupled reduced matrix elements in length $L$ and
velocity $V$ forms for transitions between the selected odd-parity
core-excited states with $J$ = 1/2 and the ground $4s$ and
singly excited $4d_{3/2}$ states in W$^{45+}$ ion.}
\begin{ruledtabular}\begin{tabular}{lrrrrrrrr}
\multicolumn{1}{c}{$4lj4l'j'(J_1)3l''j''$}&
\multicolumn{1}{c}{$Z^{(1)}_L$} & \multicolumn{1}{c}{$Z^{(1)}_V$}
& \multicolumn{1}{c}{$Z^{(2)}_L$} &
\multicolumn{1}{c}{$Z^{(2)}_V$} & \multicolumn{1}{c}{$B^{(2)}_L$}
& \multicolumn{1}{c}{$B^{(2)}_V$} & \multicolumn{1}{c}{$P^{(\rm
derv)}_L$} &
\multicolumn{1}{c}{$P^{(\rm derv)}_V$}\\[0.25pc]
\hline
\multicolumn{9}{c} {[$4lj4l'j'(J_1)3l''j''\ (1/2)$-- $4s_{1/2}$] transitions} \\
\hline
$4s_{1/2}4p_{1/2}(1)3d_{3/2}$& 0.047738& 0.044870& 0.002168&-0.000123&-0.000050& 0.001693& 0.047424& 0.000014\\
$4s_{1/2}4p_{3/2}(2)3d_{5/2}$& 0.052275& 0.049040& 0.003284& 0.000010&-0.000034& 0.002425& 0.051930&-0.000079\\
$4s_{1/2}4f_{5/2}(2)3d_{5/2}$& 0.041125& 0.038959&-0.001778& 0.000268& 0.000087&-0.000826& 0.041123& 0.000270\\
$4s_{1/2}4f_{5/2}(3)3d_{5/2}$& 0.034757& 0.032927&-0.001517& 0.000227& 0.000074&-0.000842& 0.034756& 0.000225\\
$4s_{1/2}4d_{3/2}(1)3p_{1/2}$&-0.067725&-0.064364& 0.000594& 0.000439&-0.000594& 0.000216&-0.066887&-0.000685\\
$4s_{1/2}4p_{3/2}(1)3s_{1/2}$&-0.043751&-0.041541&-0.017203& 0.000579&-0.000761&-0.015917&-0.042845&-0.000890\\
\hline
\multicolumn{9}{c} {[$4lj4l'j'(J_1)3l''j''\ (1/2)$-- $4d_{3/2}$] transitions} \\
\hline
$4p_{1/2}4d_{3/2}(2)3d_{3/2}$& 0.046223& 0.043446& 0.002084&-0.000119&-0.000054& 0.001629& 0.045918& 0.000010\\
$4p_{3/2}4d_{3/2}(3)3d_{5/2}$& 0.048898& 0.045873& 0.002691& 0.000010&-0.000028& 0.002093& 0.048576&-0.000065\\
$4s_{1/2}4d_{3/2}(2)3p_{3/2}$& 0.062272& 0.058739& 0.001399&-0.000036& 0.000201& 0.001391& 0.061840& 0.000223\\
$4d_{3/2}4f_{7/2}(2)3d_{5/2}$& 0.189383& 0.179458&-0.008551&-0.001590& 0.000298&-0.003992& 0.187961& 0.000470\\
$4d_{3/2}4f_{7/2}(3)3d_{5/2}$&-0.146695&-0.139008& 0.006235& 0.001231&-0.000236& 0.003225&-0.145594&-0.000376\\
$4d_{3/2}4f_{5/2}(1)3d_{3/2}$& 0.146008& 0.138475&-0.007222&-0.000623& 0.000324&-0.003488& 0.145197& 0.000535\\
\end{tabular}\end{ruledtabular}
\label{tab-uncl}
\end{table*}

\begin{table}
\caption{Line strengths (a.u.) calculated in length $L$ and
velocity $V$ forms for transitions between the selected odd-parity
core-excited states with $J$ = 1/2 and the ground $4s$ and
singly excited $4d_{3/2}$ states in W$^{45+}$ ion.}
\begin{ruledtabular}\begin{tabular}{lrrrrrrrr}
\multicolumn{1}{c}{}&
\multicolumn{2}{c}{First order}& \multicolumn{2}{c}{RMBPT}\\
\multicolumn{1}{c}{Upper level}& \multicolumn{1}{c}{$L$} &
\multicolumn{1}{c}{$V$} & \multicolumn{1}{c}{$L$} &
\multicolumn{1}{c}{$V$} \\[0.25pc]
\hline
\multicolumn{5}{c} {[$4lj4l'j'(J_1)3l''j''\ (1/2)$-- $4s_{1/2}$] transitions} \\
\hline
$4s_{1/2}4p_{1/2}(1)3d_{3/2}$& 0.002056& 0.001818& 0.002152& 0.002151\\
$4p_{3/2}4d_{3/2}(2)3d_{5/2}$& 0.019622& 0.017609& 0.017775& 0.017794\\
$4s_{1/2}4f_{5/2}(2)3d_{3/2}$& 0.004205& 0.003780& 0.003862& 0.003865\\
$4p_{3/2}4d_{3/2}(2)3d_{3/2}$& 0.071111& 0.063933& 0.065253& 0.065476\\
$4s_{1/2}4d_{5/2}(2)3p_{3/2}$& 0.009202& 0.008321& 0.009124& 0.009165\\
$4p_{3/2}4p_{3/2}(2)3p_{3/2}$& 0.001041& 0.000942& 0.001092& 0.001093\\
\hline
\multicolumn{5}{c} {[$4lj4l'j'(J_1)3l''j''\ (1/2)$-- $4d_{3/2}$] transitions} \\
\hline
$4s_{1/2}4d_{3/2}(2)3p_{3/2}$& 0.003858& 0.003437& 0.004065& 0.004078\\
$4d_{3/2}4f_{7/2}(3)3d_{5/2}$& 0.006774& 0.006079& 0.006224& 0.006230\\
$4d_{5/2}4f_{5/2}(3)3d_{5/2}$& 0.014631& 0.013129& 0.013505& 0.013423\\
$4p_{1/2}4p_{1/2}(0)3p_{1/2}$& 0.003732& 0.003355& 0.003401& 0.003407\\
$4d_{5/2}4f_{5/2}(1)3d_{3/2}$& 0.057817& 0.051983& 0.052786& 0.052932\\
$4d_{5/2}4f_{7/2}(2)3d_{3/2}$& 0.006813& 0.006127& 0.006215& 0.006210\\
\end{tabular}
\end{ruledtabular}
\label{tab-str}
\end{table}

\section{Method}
Details of the RMBPT  method were presented in Ref.~\cite{na-en}
for calculation of energies of particle-particle-hole states  and
in Ref.~\cite{na-tr} for calculation of radiative transition rates
from particle-particle-hole state to one-particle state.
Differences between calculations for Na-like and Cu-like ions are
due to the increased size of the model space ($4l'4l''3l^{-1}$
instead of $3l'3l''2l^{-1}$) and differences in  the Dirac-Fock
potential ($1s^22s^22p^63s^23p^63d^{10}$ instead of
$1s^22s^22p^6$), leading to 1134 states instead of 106 and more
laborious numerical calculations.

As a first step, we determine and store the single-particle
contributions to the energies for five $n$=3 hole states ($3s$,
$3p_{1/2}$, $3p_{3/2}$, $3d_{3/2}$, and $3d_{5/2}$) and   the
seven $n$=4 valence states ($4s$, $4p_{1/2}$, $4p_{3/2}$,
$4d_{3/2}$, $4d_{5/2}$, $4f_{5/2}$, and $4f_{7/2}$) in lowest,
first, and second orders.
 Next, we
evaluate and store the 664 two-particle  $\langle 4l 4l'\: J |
H^{\rm eff} |4l''4l''' \: J \rangle$ matrix elements  and the 1127
hole-particle  $\langle 3l 4l'\: J | H^{\rm eff} |3l''4l''' \: J
\rangle$ matrix elements of the effective Hamiltonian in  first
and second orders. It should be noted that these one-particle,
two-particle, and hole-particle matrix elements were used
previously to evaluate energies of the $4l4l'$ levels in Zn-like
ions \cite{saf1} and energies of the $3l^{-1}4l'$ levels in
Ni-like ions \cite{ni}.  Finally, second-order
particle-particle-hole matrix elements are evaluated \cite{na-en}.
Combining these data using the method described below, we calculate
one-, two-, and three-body contributions to the energies of
Cu-like ions.

The calculations are carried out using  sets of basis
Dirac-Hartree-Fock (DHF) orbitals. The orbitals used in the
present calculation are obtained as linear combinations of
B-splines. These B-spline basis orbitals are determined using the
method described in Ref.~\cite{2wrj}. Forty B-splines of order
of eight for each single-particle angular momentum state are used and
all orbitals with orbital angular momentum $l \leq 7$ are included in the basis
set.

\subsection{Model space}
The model space for core-excited $4l4l'3l^{-1}$  states of
copperlike ions includes 563 odd-parity states consisting of 78
$J$=1/2 states, 131 $J$=3/2 states, 143 $J$=5/2 states, 125
$J$=7/2 states, and 86 $J$=9/2 states. Additionally, there are 571
even-parity states consisting of  78 $J$=1/2 states, 131 $J$=3/2
states, 148 $J$=5/2 states, 125 $J$=7/2 states, and 89 $J$=9/2
states.  The distribution of some of the 1134 states in the model
space is summarized in Table I of the accompanying EPAPS document
\cite{EPAPS}.
% Instead of using the $4l'4l''[J_{1}]3l^{-1}(J)$ designations,
% we use  simpler
%designations $4l'4l''[J_{1}]3l(J)$ in this table. The same
%($4l'4l''3l$) designations are used for simplicity in all
%following tables and the text  below.

\subsection{Energy-matrix elements }

The evaluation of the second-order energies for the
$4l4l'(J_{1})3l''\ (J)$ states in Cu-like ions follows the pattern
of the corresponding calculation for Zn-like and Ni-like ions
given in Refs.~\cite{saf1,ni}. In particular, we use the
second-order one- and two-particle matrix elements for Zn-like
ions calculated in \cite{saf1} and hole-particle matrix elements
for Ni-like ions calculated in \cite{ni}, but recoupled as
described below, to obtain the one- and two-particle contributions
for Cu-like ions.
 We will discuss how these matrix
elements are combined to obtain the one- and two-particle
contributions to energies of Cu-like ions. We refer the reader to
Ref.~\cite{saf1,ni} for a discussion of the how the basic one- and
two-particle matrix elements were evaluated. An intrinsic
particle-particle-hole diagram  also contributes to the
second-order energy for Cu-like ions. It should be noted that the
angular part of the expression for the particle-particle-hole
diagram differs from the corresponding expression for the
three-particle matrix elements given in Ref.~\cite{bor2}. A
detailed discussion of this difference is given in
Ref.~\cite{na-en}.

Table~\ref{tab-e2} (see also  Table II of  \cite{EPAPS})
provides an illustration of various contributions to the second-order energies for the
special case of Cu-like tungsten, $Z$ = 74. In this table, we show
the one-body, two-body and three-body second-order Coulomb
contributions to the energy matrix labelled as $E^{(2)}_{i}$, $i$ = 1, 2, 3,
 The one-body second-order energy, $E^{(2)}_{1}$  is
obtained as the  sum  of the two valence and one hole $E^{(2)}_v$
energies.  The values of $E^{(2)}_{1}$  are non-zero only for
diagonal matrix elements. Even for odd-parity states with $J$=1/2
there are 78 diagonal and 6006 non-diagonal matrix elements for
$4l4l'(J_{1})3l''\ (J)$ core-excited  states.
 We list data only for the first ten diagonal matrix elements
 of odd-parity states with $J$=1/2 in
Table~\ref{tab-e2} (a more complete set of data is given in  Table
II of  \cite{EPAPS}).  It can be seen from the table that two-body
and three-body second-order contributions are positive, when the
one-body contributions are negative.
 The three-body contributions give about 20\% in the
total second-order contributions.  The values of
the $E^{(2)}_{3}$ and  $E^{(2)}_{2}$  non-diagonal contributions
are smaller than values of diagonal contributions by factor of 3-5.

After evaluating the energy matrices,  eigenvalues and eigenvectors
are calculated  for states with given values of $J$ and parity.
There are two possible methods  to  carry out the diagonalization:
(a) diagonalize the sum of  zeroth-  and  first-order  matrices,
then calculate the second-order contributions using the resulting
eigenvectors; or (b) diagonalize the  sum of  the zeroth-,  first-
and  second-order matrices together. Following Ref.~\cite{bor2},
we choose the second method here.

Table \ref{tab-e4} lists the energies of ten
excited states of W$^{45+}$ from Table \ref{tab-e2}
including the total energies $E_{\rm {tot}}$. The latter is the sum of
the following contributions: $E^{(0+1)}$ =
$E^{(0)}$ +$E^{(1)}$+$B^{(1)}$, the second-order Coulomb and Breit
energy $E^{(2)}$, and the QED correction $E_{\rm {LAMB}}$. The QED correction is
approximated as the sum of the one-electron self energy and the
first-order vacuum-polarization energy. The screened self-energy
and vacuum polarization data given by Kim {\it et al.\/}
\cite{kim} are used to determine the QED  correction $E_{\rm
{LAMB}}$ (see, for detail Ref.~\cite{na-en}). As can be seen,
the second order correction contributes  to the total  energy
from 0.4 \%  for lowest levels up to 8 \% for high-excited levels.
The levels in this table (see also  Table III of
\cite{EPAPS})  could be divided into groups corresponding to
excited $4l4l'$ states and $3l$ hole states.

\begin{figure}[tbp]
\centerline{\includegraphics[scale=0.35]{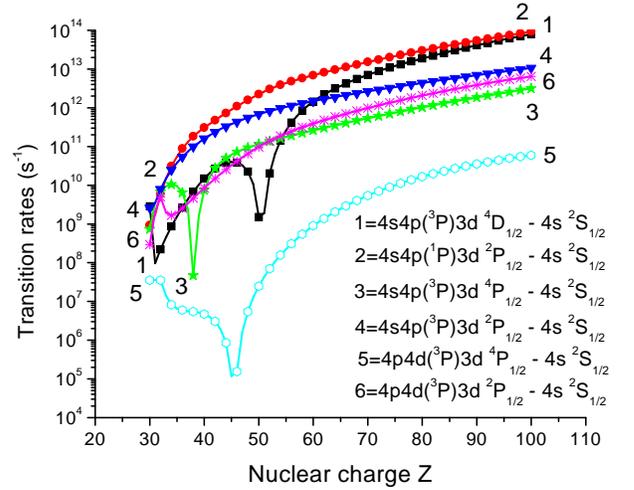}}
\caption{Weighted transition rates between core-excited odd-parity
states with $J$ = 1/2 and $4s\ ^2S_{1/2}$ states as function of
$Z$ in Cu-like ions.} \label{od1}
\end{figure}

\begin{figure}[tbp]
\centerline{\includegraphics[scale=0.4]{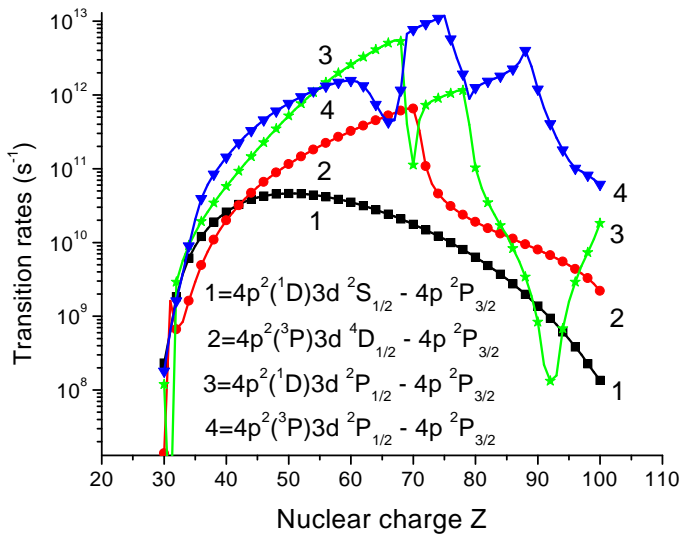}}
\caption{Weighted transition rates between core-excited
even-parity states with $J$ = 1/2 and  $4p\ ^2P_{3/2}$ states as
function of $Z$ in Cu-like ions. } \label{ev1}
\end{figure}

\begin{table*}
\caption{Wavelengths ($\lambda$ in \AA) and transition rates
($A_r$ in $s^{-1}$)  for transitions from core-excited states $QJ$
($Q = 4lj4l'j'(J_1)3l''j''$) to the the ground $4s$ and
 singly excited $4p_{1/2}$ states in W$^{45+}$ ion. Comparison with
theoretical data obtained by using {\sc cowan} code from
Ref.~\protect\cite{cowan}.
 Numbers in brackets represent powers of 10.}
\begin{ruledtabular}\begin{tabular}{lrrrrrl}
\multicolumn{1}{c} {$jj$ coupling} & \multicolumn{1}{c} {$\lambda_{\rm
MCDF}$}& \multicolumn{1}{c} {$\lambda_{\rm RMBPT}$}&
\multicolumn{1}{c} {$A_{r}^{\rm RMBPT}$}& \multicolumn{1}{c}
{$\lambda_{\sc cowan}$}&
\multicolumn{1}{c} {$A_{r}^{\sc cowan}$}&
 {$LS$ coupling}\\
\hline
\multicolumn{6}{c} {[$4lj4l'j'(J_1)3l''j''$-- $4s_{1/2}$] transitions} \\
$4p_{3/2}4d_{3/2}(1)3d_{5/2}$& 5.919& 5.9248& 2.763[14]& 5.9110& 3.020[14]& $4s4f3d(^3D)\ ^4D_{3/2}$\\
$4p_{3/2}4d_{3/2}(2)3d_{5/2}$& 5.912& 5.9200& 1.735[14]& 5.9069& 1.786[14]& $4s4f3d(^3D)\ ^4D_{1/2}$\\
$4s_{1/2}4f_{7/2}(3)3d_{3/2}$& 5.752& 5.7589& 1.863[14]& 5.7578& 1.832[14]& $4p4d3d(^3P)\ ^4F_{3/2}$\\
$4p_{3/2}4d_{3/2}(0)3d_{3/2}$& 5.744& 5.7512& 2.824[14]& 5.7501& 3.295[14]& $4p4d3d(^3P)\ ^4F_{3/2}$\\
$4p_{3/2}4d_{3/2}(2)3d_{3/2}$& 5.718& 5.7291& 7.026[14]& 5.7274& 7.954[14]& $4s4f3d(^3D)\ ^2P_{1/2}$\\
$4p_{3/2}4d_{3/2}(2)3d_{3/2}$& 5.715& 5.7239& 8.195[14]& 5.7243& 9.738[14]& $4s4f3d(^3D)\ ^2P_{3/2}$\\
$4p_{1/2}4p_{1/2}(0)3p_{3/2}$& 5.615& 5.6347& 9.915[13]& 5.6419& 1.003[14]& $4p4d3d(^3P)\ ^2P_{3/2}$\\
$4s_{1/2}4d_{5/2}(2)3p_{3/2}$& 5.230& 5.2409& 3.212[14]& 5.2322& 3.023[14]& $4s4d3p(^1P)\ ^2P_{3/2}$\\
$4s_{1/2}4d_{5/2}(2)3p_{3/2}$& 5.222& 5.2328& 1.289[14]& 5.2252& 1.288[14]& $4s4d3p(^3P)\ ^2P_{1/2}$\\
\multicolumn{6}{c} {[$4lj4l'j'(J_1)3l''j''$-- $4p_{1/2}$] transitions} \\
$4s_{1/2}4p_{3/2}(1)3p_{3/2}$& 5.912& 5.9175& 4.245[14]& 5.9093& 5.406[14]& $4p4f3d(^3F)\ ^4F_{3/2}$\\
$4s_{1/2}4p_{3/2}(1)3p_{3/2}$& 5.910& 5.9179& 2.314[14]& 5.9074& 2.855[14]& $4p4f3d(^3F)\ ^4D_{1/2}$\\
$4p_{1/2}4f_{7/2}(3)3d_{3/2}$& 5.752& 5.7544& 5.714[13]& 5.7407& 1.497[14]& $4d4d3d(^1G)\ ^2D_{3/2}$\\
$4d_{3/2}4d_{5/2}(1)3d_{5/2}$& 5.732& 5.7349& 1.680[14]& 5.7206& 6.059[14]& $4d4d3d(^3P)\ ^2D_{3/2}$\\
$4d_{3/2}4d_{5/2}(2)3d_{5/2}$& 5.723& 5.7325& 1.235[15]& 5.7256& 8.290[14]& $4p4f3d(^1D)\ ^2D_{3/2}$\\
$4d_{3/2}4d_{5/2}(2)3d_{5/2}$& 5.722& 5.7267& 4.256[14]& 5.7210& 7.570[14]& $4p4f3d(^3F)\ ^2P_{1/2}$\\
$4d_{3/2}4d_{5/2}(3)3d_{5/2}$& 5.719& 5.7268& 2.074[14]& 5.7130& 4.057[13]& $4p4f3d(^1D)\ ^2P_{1/2}$\\
$4p_{1/2}4d_{5/2}(2)3p_{3/2}$& 5.233& 5.2431& 1.086[14]& 5.2354& 7.306[13]& $4p4d3p(^3S)\ ^2D_{3/2}$\\
$4p_{1/2}4d_{5/2}(3)3p_{3/2}$& 5.230& 5.2400& 2.323[14]& 5.2548& 2.845[14]& $4p4d3p(^3D)\ ^2D_{3/2}$\\
$4p_{1/2}4d_{5/2}(2)3p_{3/2}$& 5.229& 5.2395& 1.825[14]& 5.2546& 1.680[14]& $4p4d3p(^3D)\ ^2P_{1/2}$\\
\end{tabular}
\end{ruledtabular}
\label{tab-cowan1}
\end{table*}

\begin{table}
\caption{Wavelengths ($\lambda$ in \AA) and transition rates ($A_r$ in $s^{-1}$)  for transitions from
core-excited states $QJ$ ($Q = 4lj4l'j'(J_1)3l''j''$) to the
the ground $4s$ and
singly excited $4l_{j}$ states in Cu-like Ce, W, and Au ions.
Comparison  of theoretical data obtained by using RMBPT  code with experimental measurements
 ($\lambda_{\rm expt}$) from
Ref.~\protect\cite{doron-scr} ($a$), Ref.~\protect\cite{klapisch}
($b$), and Ref.~\protect\cite{luc} ($c$)). Numbers in brackets
represent powers of 10.}
\begin{ruledtabular}\begin{tabular}{rrrrr}
\multicolumn{2}{c} {Transition} & \multicolumn{1}{c}
{$\lambda_{\rm RMBPT}$}& \multicolumn{1}{c} {$A_{r}^{\rm RMBPT}$}&
\multicolumn{1}{c} {$\lambda_{\rm expt}$}\\
\hline
                       \multicolumn{4}{c} {Cu-like Ce, $Z$=58} \\
$4s_{1/2}4f_{5/2}(3) 3p_{1/2}(5/2)$&$ 4p_{3/2}$&   9.4558& 4.16[13]&  9.452$^a$\\
$4s_{1/2}4p_{3/2}(2) 3s_{1/2}(5/2)$&$ 4d_{3/2}$&   9.9569& 8.32[12]&  9.940$^a$\\
$4d_{5/2}4d_{5/2}(2) 3p_{3/2}(5/2)$&$ 4d_{5/2}$&  10.0518& 1.58[13]& 10.051$^a$\\
$4p_{3/2}4d_{3/2}(2) 3p_{3/2}(1/2)$&$ 4p_{3/2}$&  10.1114& 1.96[13]& 10.111$^a$\\
$4d_{3/2}4d_{5/2}(4) 3p_{3/2}(5/2)$&$ 4d_{3/2}$&  10.1435& 3.52[13]& 10.141$^a$\\
$4p_{3/2}4f_{7/2}(2) 3d_{3/2}(3/2)$&$ 4p_{1/2}$&  11.3223& 8.62[11]& 11.321$^a$\\
$4p_{1/2}4p_{3/2}(2) 3p_{1/2}(5/2)$&$ 4d_{3/2}$&  11.4202& 7.77[12]& 11.412$^a$\\
$4d_{3/2}4d_{5/2}(1) 3d_{3/2}(1/2)$&$ 4p_{1/2}$&  11.4912& 1.24[14]& 11.491$^a$\\
$4d_{5/2}4f_{5/2}(3) 3d_{5/2}(3/2)$&$ 4d_{3/2}$&  11.8120& 1.64[13]& 11.816$^a$\\
                   \multicolumn{4}{c} {Cu-like W, $Z$=74} \\
$4p_{3/2}4d_{5/2}(3) 3d_{3/2}(5/2)$&$ 4d_{5/2}$&   6.8197& 2.83[12]&  6.816$^b$\\
$4p_{3/2}4f_{7/2}(3) 3d_{3/2}(5/2)$&$ 4f_{7/2}$&   6.8269& 4.26[12]&  6.827$^b$\\
$4p_{1/2}4p_{3/2}(2) 3d_{3/2}(1/2)$&$ 4p_{1/2}$&   6.8577& 2.32[12]&  6.858$^b$\\
$4s_{1/2}4p_{3/2}(1) 3d_{3/2}(3/2)$&$ 4s_{1/2}$&   6.8885& 3.93[12]&  6.884$^b$\\
$4d_{3/2}4d_{5/2}(4) 3d_{3/2}(5/2)$&$ 4f_{7/2}$&   6.8947& 1.89[12]&  6.896$^b$\\
$4p_{3/2}4d_{5/2}(3) 3d_{5/2}(7/2)$&$ 4d_{5/2}$&   7.0724& 3.41[13]&  7.075$^b$\\
$4p_{3/2}4f_{7/2}(3) 3d_{5/2}(5/2)$&$ 4f_{7/2}$&   7.1095& 1.31[12]&  7.108$^b$\\
$4p_{3/2}4f_{7/2}(3) 3d_{5/2}(9/2)$&$ 4f_{7/2}$&   7.1117& 6.78[12]&  7.113$^b$\\
$4p_{3/2}4f_{5/2}(3) 3d_{5/2}(5/2)$&$ 4f_{7/2}$&   7.1302& 1.96[13]&  7.131$^b$\\
$4p_{3/2}4d_{5/2}(1) 3d_{5/2}(7/2)$&$ 4d_{5/2}$&   7.1354& 1.59[13]&  7.137$^b$\\
$4s_{1/2}4f_{7/2}(4) 3d_{5/2}(5/2)$&$ 4d_{5/2}$&   7.2396& 1.61[12]&  7.242$^b$\\
$4s_{1/2}4d_{5/2}(2) 3d_{5/2}(3/2)$&$ 4p_{3/2}$&   7.2479& 2.06[13]&  7.248$^b$\\
$4s_{1/2}4f_{7/2}(3) 3d_{5/2}(3/2)$&$ 4d_{5/2}$&   7.2608& 1.58[13]&  7.262$^b$\\
$4p_{1/2}4d_{3/2}(1) 3d_{3/2}(5/2)$&$ 4d_{3/2}$&   7.2916& 1.64[13]&  7.293$^b$\\
                   \multicolumn{4}{c} {Cu-like Au, $Z$=79} \\
$4d_{5/2}4d_{5/2}(4) 3p_{3/2}(7/2)$&$ 4d_{5/2}$&   4.3974& 1.65[14]&   4.39$^c$\\
$4d_{3/2}4f_{7/2}(4) 3p_{3/2}(9/2)$&$ 4f_{7/2}$&   4.4690& 1.79[14]&   4.69$^c$\\
$4d_{5/2}4f_{7/2}(2) 3d_{3/2}(5/2)$&$ 4d_{3/2}$&   4.7690& 1.66[14]&   4.76$^c$\\
$4f_{7/2}4f_{7/2}(4) 3d_{3/2}(7/2)$&$ 4f_{7/2}$&   4.7997& 4.53[15]&   4.80$^c$\\
$4f_{5/2}4f_{7/2}(4) 3d_{3/2}(7/2)$&$ 4f_{5/2}$&   4.8214& 1.98[15]&   4.82$^c$\\
$4d_{5/2}4d_{5/2}(4) 3d_{5/2}(5/2)$&$ 4p_{3/2}$&   4.9373& 2.51[14]&   4.93$^c$\\
$4d_{5/2}4f_{5/2}(3) 3d_{5/2}(3/2)$&$ 4d_{3/2}$&   4.9610& 2.58[14]&   4.96$^c$\\
$4s_{1/2}4p_{3/2}(2) 3p_{3/2}(1/2)$&$ 4p_{3/2}$&   5.2208& 1.15[13]&   5.22$^c$\\
$4d_{3/2}4d_{5/2}(3) 3d_{3/2}(7/2)$&$ 4f_{5/2}$&   5.7207& 2.08[12]&   5.72$^c$\\
$4s_{1/2}4p_{3/2}(2) 3d_{5/2}(3/2)$&$ 4s_{1/2}$&   5.9029& 3.93[13]&   5.90$^c$\\
$4p_{3/2}4d_{5/2}(1) 3d_{5/2}(7/2)$&$ 4d_{5/2}$&   5.9427& 2.10[13]&   5.94$^c$\\
$4p_{3/2}4d_{3/2}(2) 3d_{5/2}(7/2)$&$ 4d_{5/2}$&   6.0107& 1.28[12]&   6.01$^c$\\
$4s_{1/2}4d_{5/2}(3) 3d_{5/2}(1/2)$&$ 4p_{3/2}$&   6.0801& 1.61[13]&   6.08$^c$\\
\end{tabular}
\end{ruledtabular}
\label{tab-expt}
\end{table}

\subsection{Dipole matrix element}

We designate the first-order dipole matrix element by $Z^{(1)}$,
the Coulomb-correction to the second-order matrix element
$Z^{(2)}$, and the second-order Breit correction $B^{(2)}$. The
evaluation of $Z^{(1)}$, $Z^{(2)}$, and $B^{(2)}$ for Cu-like ions
follows the pattern of the corresponding calculation for Na-like
ions in Ref.~\cite{na-tr}. These matrix elements are calculated in
both length and velocity gauges.

 Table~\ref{tab-uncl} lists values of uncoupled first-and second-order
dipole matrix elements $Z^{(1)}$, $Z^{(2)}$,
$B^{(2)}$, together with derivative terms $P^{({\rm derv})}$ for
Cu-like tungsten, $Z$ = 74. For simplicity, only the values
for the selected dipole transitions between  odd-parity  states
with $J$ = 1/2 and the ground $4s$ and excited $4d_{3/2}$ states are presented.
The more comprehensive set of data is given  in Table IV of \cite{EPAPS}).
The derivative terms shown in Table~\ref{tab-uncl} arise because
transition amplitudes depend on energy, and the transition energy
changes order-by-order in RMBPT calculations. Both length ($L$)
and velocity ($V$) forms are given for the matrix elements. We
find that the first-order matrix elements $Z^{(1)}_L$ and
$Z^{(1)}_V$ differ by 10\%; the $L$ - $V$ differences between
second-order matrix elements are much larger for some transitions.
The term  $P^{({\rm derv)}}$ in length form almost equals
$Z^{(1)}$ in length form but
 in velocity form is smaller than $Z^{(1)}$
 in length form by three to four orders of magnitude.

Values of line strengths calculated as a square of coupled reduced
matrix element \cite{na-tr} in length and velocity forms are given
in Table \ref{tab-str} for the  selected dipole transitions
between  odd-parity  states with $J$ = 1/2 and the ground $4s$ and
excited $4d_{3/2}$ states. A more complete set of data is given
Table V of \cite{EPAPS}. Although we use an intermediate-coupling
scheme, it is nevertheless convenient to label the physical states
using the $jj$ scheme. We see that $L$ and $V$ forms of the
coupled matrix elements in Table~\ref{tab-str} differ  only in the
third or fourth digits. These $L$--$V$ differences arise because
 our RMBPT calculations start with a non-local Dirac-Fock (DF)
potential.  If we were to replace the DF potential by a local
potential, the differences would disappear completely. The first
two columns in Table \ref{tab-str} show $L$ and $V$ values of
line strengths calculated in the first-order approximation which differ
by factor of 10. Thr last two columns indicate that including of the second-order
contribution almost removes the $L-V$ differences.

\section{Results and comparisons with other theory and experiment}
We calculate energies of  core-excited even-parity $4lj4l'j'(J_1)3l''j''$
 as well as  the  singly excited $4lj$ states in Cu-like ions with $Z$ = 30 -100.
  Reduced matrix elements, line strengths, oscillator
strengths, and transition rates are also determined for
electric dipole transitions between the above mentioned  core-excited
 and   singly excited states in Cu-like ions for the same range of  $Z$.

Table~\ref{tab-cowan1} lists theoretical data for
selected transitions from core-excited levels to the ground
$4s$ and a singly excited $4p_{1/2}$ levels with largest values of transition rates in
 W$^{45+}$. Also, the comparison of theoretical data produced
 by the different methods and codes is inlcuded. Specifically, three values for
wavelengths:
 $\lambda_{\rm MCDF}$, $\lambda_{\rm RMBPT}$, and $\lambda_{\sc cowan}$
are compared. The first values,
 $\lambda_{\rm MCDF}$, were obtained as the first-order RMBPT values.
It can be seen from  Table~\ref{tab-cowan1} that the results
 obtained by {\sc cowan} code~\cite{cowan}, $\lambda_{\sc cowan}$, better agree with
 the MCDF results, $\lambda_{\rm MCDF}$ than with
 the RMBPT results, $\lambda_{\rm RMBPT}$.

{\sc cowan} code \cite{cowan} gives results
which are generally in good agreement with experimental energies
by scaling the electrostatic Slater parameters to include the
correlation effects (Refs.~\cite{sataka})
and here the scaling factor of 0.85 was used.
Also, in {\sc cowan} code the $LS$ coupling scheme is
implemented and corresponding $LS$ designations were added to $jj$
designations in all tables which include comparisons
with theoretical data produced by  {\sc cowan} code.
%We  mentioned previously neither $jj$ nor $LS$ coupling can
%describe the physical states
% properly but more often  the $LS$ designations are chosen to
%compare with available experimental data, especially, for low-$Z$ ions.
% The $jj$
%designations are used in RMBPT code as initial to calculate the energy matrix. These
%primary designations do not coincide with the $jj$ designations
%obtained by largest mixing coefficients.  Using $jj$ designations by the
%largest mixing coefficients leads to crossing of levels. To avoid
%crossing of levels, we do not follow the $jj$ label of the largest
%mixing coefficient and  write the $jj$ label used as initial for
%%noncouple   matrix elements. Sometimes, this is not convenient
%since it leads to the difference in designations chosen for
%small-$Z$ ions. In {\sc cowan} code, the $LS$-labels are chosen by largest mixing coefficients.
%This choice leads to use the same label for two levels with different energies.
% We used a different coupling in RMBPT ($jj$
%coupling) and {\sc cowan} code ($LS$ coupling) and an agreement of both
%results confirm our identification of levels when sometimes even
%configurations are different.

A comprehensive set of theoretical data and comparisons
for transitions with largest values of transition rates in
 W$^{45+}$ similar to presented in Table~\ref{tab-cowan1}
is given in Table VI of the accompanying EPAPS document \cite{EPAPS}.
It includes transitions from core-excited levels not only to the ground $4s$
and to one singly excited level  $4p_1/2$ as in Table~\ref{tab-cowan1} but
to all singly excited $4l_{j}$ levels.

The similar comprehensive set of theoretical data and
comparisons for transitions between core-excited states
$4lj4l'j'(J_{1}) 3l''\ J$ to the ground $4s$ and
singly excited $4l_{j}$ states is presented for Ce$^{29+}$
ion in Table VII of Ref.~\cite{EPAPS} and for Au$^{50+}$ ion in Table VIII of Ref.~\cite{EPAPS}.
Comparisons with theoretical data obtained by using
RMBPT, MCDF, and {\sc cowan} codes show that
 the difference in results can be
 explained by the second-order corrections to energy and dipole
 matrix elements included in RMBPT.

The trends of the $Z$-dependence of transition rates for the
transitions from core-excited odd-parity states with $J$ = 1/2 to
the ground state $4s\ ^2S_{1/2}$ states  are presented in
Fig.~\ref{od1} . The trends for the transitions from core-excited
even-parity states with $J$ = 1/2 to the singly excited state $4p\
^2P_{3/2}$   are shown in Fig.~\ref{ev1}. More transition rate
figures are given in \cite{EPAPS}.

 We see from the graphs that transitions with smooth $Z$-dependence
are rarer than transitions with sharp features.  Smooth
 $Z$-dependence occurs for  transitions from doublet and quartet
 core-excited states. Usually, singularities  happen in the
 intermediate interval of $Z$ = 40 - 60 when neither $LS$ nor $jj$ coupling
schemes describe the states of these ions properly.
  One general conclusion
 that can be derived from these graphs is that the smooth
$Z$-dependence occurs more frequently for transitions from
low-lying core-excited states.

Singularities in the transition-rate curves have two distinct
origins: avoided level crossings and zeros in dipole matrix
elements. Avoided level crossings result in changes of the
dominant configuration of a state at a particular value of $Z$ and
lead to abrupt changes in the transition rate curves when the
partial rates associated with the dominant configurations below
and above the crossing point are significantly different. Zeros in
transition matrix elements lead to cusp-like minima in the
transition rate curves.  Examples of each of these two singularity
types can be seen in Figs.~\ref{od1} and ~\ref{ev1}, as well as in
Figs.~1 - 10 presented in EPAPS document \cite{EPAPS}.

In Table~\ref{tab-expt}, wavelengths and electric-dipole
transition rates along with comparison with experimental data
 are presented for transitions in Cu-like Ce, W, and Au.
  The table is limited to identification of experimentally measured transitions
 given in Refs.~\cite{doron-scr,klapisch,luc}. A more comprehensive set of
theoretical data assigned to each experimental transition
is presented in Table~IX of EPAPS document
\cite{EPAPS}. We  mentioned previously that all possible
$4l_1j_14l_2j_2(J_1)3l_3j_3 (J)$ - $4l_{j}$ transitions produce
2294 spectrum lines.  These lines in Ce$^{29+}$,   W$^{45+}$, and
Au$^{50 +}$ cover the spectral regions from 6.8 -- 21.9~\AA~,
3.5 -- 9.8 ~\AA~, and 2.9 -- 8.7~\AA~, respectively. A number of
spectral lines becomes smaller by a factor of ten when we consider
transitions with largest values of transition rates, $gA_r$. The
number of transitions with $gA_r>$ 10$^{13}$~s$^{-1}$ is about 200
for Ce$^{29+}$,
and a number of transitions with $gA_r>$ 10$^{14}$ ~s$^{-1}$ is about 130
for W$^{45+}$ and about 190 for Au$^{50 +}$.
Those transitions with the largest values of $gA_r$ cover
the  spectral regions smaller by factor of 3-4 than that mentioned for all lines.
 We can see that the number of predicted spectral lines even with
the largest values of $gA_r$ still much larger than a number of experimental
lines shown in Table~\ref{tab-expt}. Also, the interval between these lines
 is about 0.01~\AA~ or less which is comparable with
 the accuracy of experimental measurements. In this case, it
could be reasonable to assign not a single transition as in Table~\ref{tab-expt}
but 3-5 transitions
 to identify the experimental peak, as it is demonstrated in Table IX of EPAPS
document \cite{EPAPS}.

\begin{figure}[tbp]
\centerline{\includegraphics[scale=0.35]{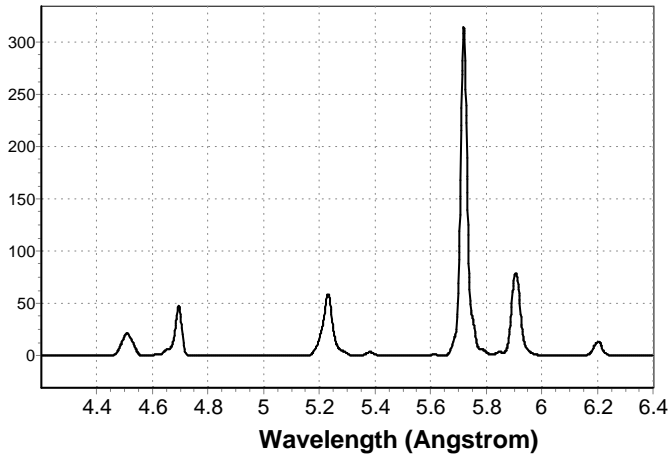}}
\caption{Synthetic spectra of Cu-like W calculated using the
 RMBPT atomic data and Gaussian profiles
 with ($\Delta \lambda$ = 0.02~Angstrom). The scale in the ordinate is in
units of 10$^{13}$~s$^{-1}$.}\label{cun-mbpt}
\end{figure}

\begin{figure}[tbp]
\centerline{\includegraphics[scale=0.35]{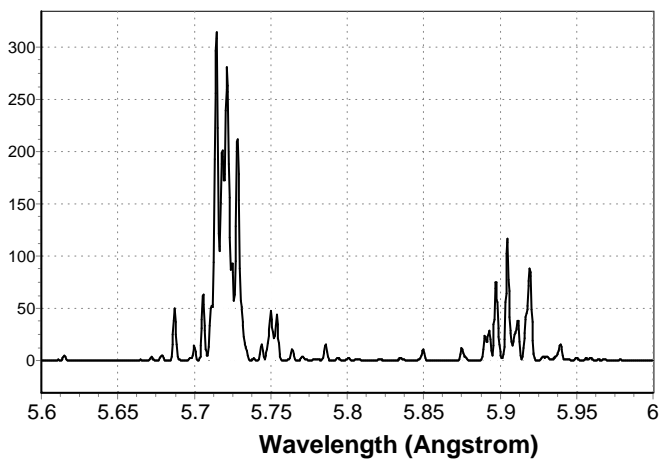}}
\caption{Synthetic spectra of Cu-like W calculated using the
 RMBPT atomic data and Gaussian profiles with
 ($\Delta \lambda$ = 0.002~Angstrom). The scale in the ordinate is in
units of 10$^{13}$~s$^{-1}$.}\label{cun1-mbpt}
\end{figure}

These atomic data are important in modeling of M-shell
radiation spectra of heavy ions generated in electron beam ion trap
experiments and in M-shell diagnostics of plasmas.
For example, x-ray M-shell spectra of W ions collected
 at the different energies of the electron beam at LLNL
  electron beam ion trap indicate the existence of strong
   Cu-like spectral features \cite{alla1}.
Also,the distinct features of x-ray M-shell spectra of W
 ions in a spectral region from 5 to 6~\AA~ produced by
 laboratory plasmas are the brightest Ni-like lines
 and Co- and Cu-like spectral features.
In particular, Cu-like satellite lines constitute  most
of the Cu-like spectral features, and Cu-like autoionization
levels make an important contribution in ionization
balance calculations \cite{alla2}.
Synthetic x-ray spectra of Cu-like W computed with
different resolution of 0.02~\AA~ and
0.002~\AA~ are presented in Fig.~\ref{cun-mbpt} ~and ~\ref{cun1-mbpt},
respectively.
It was assumed that  spectral lines have the intensities proportional
to the weighted
transition probabilities and are fitted with the Gaussian profile.
 Specifically, the spectrum in Fig.~\ref{cun-mbpt} includes
$3l-4l'$ transitions and covers the spectral region from 4 \AA~ to 6.4 \AA.
The most intense peaks at about 5.72 \AA~ and 5.9 \AA~ are
formed by $3d-4f$ transitions and
are shown with better resolution in Fig.~\ref{cun1-mbpt}.

\section{Conclusion}
In summary,  a systematic second-order RMBPT study of the energies
and transition rates for [$4l_1j_14l_2j_2(J_1)3l_3j_3\ (J)$ -
$4lj$] electric-dipole transitions in Cu-like  ions with the
nuclear charges $Z$ ranging from 30 to 100 has been presented.
 The retarded $E1$ matrix elements included
correlation corrections from Coulomb and Breit interactions. Both
length and velocity forms of the matrix elements were evaluated
and small differences (0.4 - 1 \%), caused by the non locality of
the starting DF potential, were found between the two forms.
Second-order RMBPT transition energies were used in the evaluation
of transition rates. These calculations were compared with other
calculations and with available experimental data. For $Z \geq
40$, we believe that the present theoretical data are more
accurate than other theoretical or experimental data for
transitions between the $4l_1j_14l_2j_2(J_1)3l_3j_3\ (J)$
core-excited  states and the $4lj$ singly excited states in
Cu-like ions. The results could be further improved by including
 third-order correlation
corrections.These calculations are presented as a theoretical
benchmark for comparison with experiment and theory.
In addition, the application of generated atomic data to modeling and interpreting
of x-ray M-shell spectra of heavy ions is discussed.

\begin{acknowledgments}
The work of W.R.J.  was supported in part by National Science
Foundation Grant No.\ PHY-01-39928. U.I.S. acknowledges partial
support by Grant No.\ B516165 from Lawrence Livermore National
Laboratory. The work of A.S. was supported by the DOE-NNSA/NV
Cooperative Agreement DE-FC08-01NV14050 and Sandia National Laboratories.
\end{acknowledgments}

%\bibliography{cu}

\end{document}